\documentclass[aps,prl,twocolumn,superscriptaddress,showpacs]{revtex4-1}
\usepackage[pdftex]{graphicx}   
\usepackage{nicefrac}
\usepackage{color}
\usepackage{subfigure}
\begin{document}

\title{Evidence for Long-Range Spin Order Instead of a Peierls Transition \newline in Si(553)-Au Chains}

\author{J. Aulbach}
\affiliation{Physikalisches Institut and R\"ontgen Center for Complex Materials Systems (RCCM), Universit\"at W\"urzburg, D-97074 W\"urzburg, Germany}
\author{J. Sch\"afer}
\affiliation{Physikalisches Institut and R\"ontgen Center for Complex Materials Systems (RCCM), Universit\"at W\"urzburg, D-97074 W\"urzburg, Germany}
\author{S.C. Erwin}
\affiliation{\mbox{Center for Computational Materials Science, Naval Research Laboratory, Washington, DC 20375, USA}}
\author{S. Meyer}
\affiliation{Physikalisches Institut and R\"ontgen Center for Complex Materials Systems (RCCM), Universit\"at W\"urzburg, D-97074 W\"urzburg, Germany}
\author{C. Loho}
\affiliation{Physikalisches Institut and R\"ontgen Center for Complex Materials Systems (RCCM), Universit\"at W\"urzburg, D-97074 W\"urzburg, Germany}
\author{J. Settelein}
\affiliation{Physikalisches Institut and R\"ontgen Center for Complex Materials Systems (RCCM), Universit\"at W\"urzburg, D-97074 W\"urzburg, Germany}
\author{R. Claessen}
\affiliation{Physikalisches Institut and R\"ontgen Center for Complex Materials Systems (RCCM), Universit\"at W\"urzburg, D-97074 W\"urzburg, Germany}



\date{\today}

\begin{abstract}

Stabilization of the Si(553) surface by Au adsorption results in two different atomically defined chain types, one of Au atoms and one of Si. At low temperature these chains develop two- and threefold periodicity, respectively, previously attributed to Peierls instabilities. Here we report evidence from scanning tunneling microscopy that rules out this interpretation. The $\times$3 superstructure of the Si chains vanishes for low tunneling bias, i.e., close the Fermi level. In addition, the Au chains remain metallic despite their period doubling. Both observations are inconsistent with a Peierls mechanism. On the contrary, our results are in excellent, detailed agreement with the Si(553)-Au ground state predicted by density-functional theory, where the $\times$2 periodicity of the Au chain is an inherent structural feature and every third Si atom is spin-polarized.
\end{abstract}


\maketitle


Atoms  can  form  chain-like  architectures  by  self-assembly  on  various  semiconductor  surfaces. Such chains have been widely studied because they may offer physical realizations of various one-dimensional (1D) electronic ground states -- Peierls instabilities (i.e., charge density waves, CDW) \cite{Johannes} or Tomonaga-Luttinger liquids \cite{NatureBlum} -- in which Coulomb interactions are dominant. An equally interesting scenario arises if the electron's spin degree of freedom is important or even dominant. For example, a proposal was made \cite{Mahan} to use atomic chains as a spin shift register, where spin encodes the information. Recent research has focused on spin alignment in 2D atom lattices on semiconductor substrates  \cite{Philipp, Li}. The fate of spin ordering in 1D chains on surfaces, however, has remained less explored experimentally. 

The variability offered by chains formed on different high index Si surfaces allows us to investigate this interplay of charge, spin, and lattice in a family of related structures. Specific representatives include the chain structures stabilized by Au on Si(557)-Au and Si(553)-Au. These systems carry Au-induced metallic electron bands as seen in photoemission \cite{Crain2004}. In Si(553)-Au -- the focus of the present work -- the situation is particularly complex. The structure as derived from x-ray diffraction \cite{Voegeli} and density-functional theory (DFT) \cite{Krawiec, Erwin} exhibits a dimerized double-strand Au chain, in contrast to the single Au row in Si(557)-Au. As an additional key characteristic of both variants, there is a second type of chain located at the terrace edge, formed by Si atoms which are arranged in a graphene-like honeycomb chain \cite{Krawiec, Erwin}.

In two seminal papers, changes in the periodicity of both types of chains upon cooling were observed by scanning tunneling microscopy (STM) \cite{Snijders2006, Ahn2005}, leading to two- and threefold patterns for the Au and the Si chain,
respectively. Moreover, from photoemission data \cite{Ahn2005} a temperature-dependent gap opening was inferred. These
observations for Si(553)-Au were interpreted as Peierls instabilities, driven by nesting in the metallic bands, and
resulting in energy gaps and periodic lattice distortions at low temperature.  

However, several pieces of evidence do not support this interpretation of two coexisting Peierls distortions. To begin with, the observed band filling \cite{Crain2003} does not match the conditions for nesting \cite{ReviewSnijders} required for the observed two- and threefold periodicities. Moreover, the bands induced by the heavy Au atoms are prone to a Rashba-type spin-orbit splitting, as recently demonstrated for Si(557)-Au \cite{Okuda2010}. In Si(553)-Au the band backfolding situation at the $\times$2 zone boundary \cite{Barke2006} likewise argues for Rashba band pairs. Therefore, the previously claimed different temperature-dependence of the supposed CDW gaps in such paired bands \cite{Ahn2005} does not seem plausible, casting doubts on the simple Peierls picture for this system.

A fundamentally different theoretical explanation emerges from DFT calculations \cite{Erwin}. In the DFT ground state, every third Si chain atom carries one fully spin-polarized electron with the localized spins forming an antiferromagnetically ordered chain. At low temperature, this leads to the $\times$3 periodicity seen in STM, which detects the associated charge. The DFT ground state also reveals a period doubling within the row of Au atoms. 

Thus this model embodies a qualitatively different origin, unrelated to Peierls physics, for the observed superstructures. Indications for density of states (DOS) corresponding to the spin model were found by scanning tunneling spectroscopy (STS) \cite{Snijders2012}. However, a recent STM report \cite{Shin2012} studying finite-length chains still maintains a CDW interpretation for this system.
In this light, a closer experimental scrutiny is needed to clearly distinguish between the competing scenarios.

\begin{figure*}[t!]
\includegraphics[width=\linewidth]{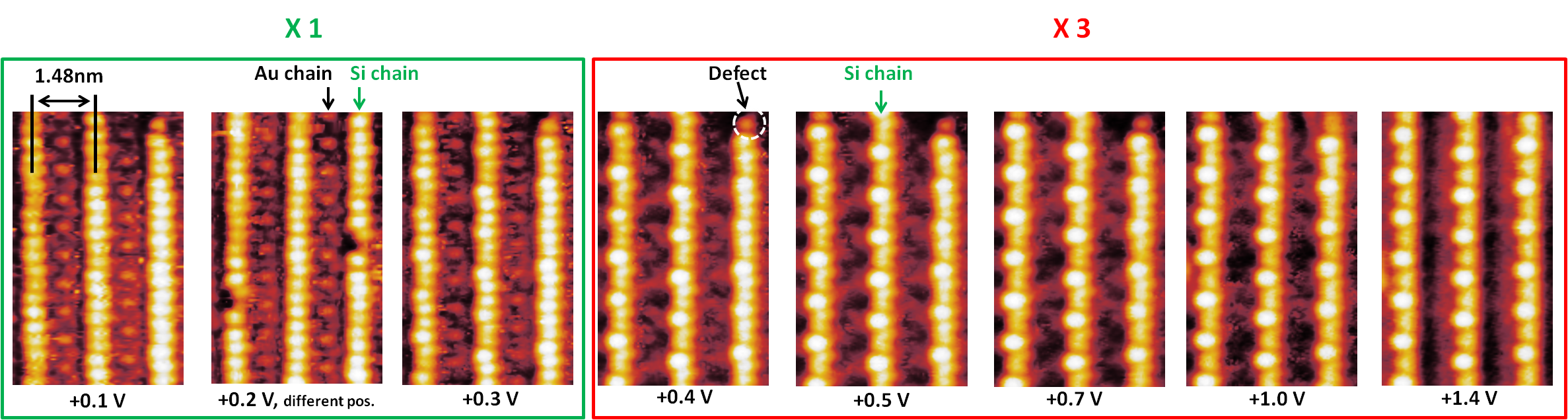}
 \caption{STM bias series of the unoccupied states ($U$~=~0.1-1.4~V, $I$~=~50~pA) of an identical sample area (except for $U$~=~0.2~V) at fixed $T$~=~77~K. For low bias the Si chain shows a $\times$1 periodicity, which changes into a $\times$3 periodicity for higher bias. This cannot be explained by a simple CDW scenario. The Au atoms exhibit a $\times$2 periodicity and are only clearly visible at low bias ($\leq$0.3~V).}
\label{fig:Biasseries}
\end{figure*}

In this Letter, we use STM, STS and DFT to definitively establish that the higher order periodicity arises not from Peierls instabilities but rather from spin-polarization of the Si chain atoms. Atomically resolved topographic measurements reveal two different periodicities ($\times$1 and $\times$3) for the Si chain at fixed temperature, depending on actual bias. This is in detailed agreement with DFT-simulated images based on the spin-polarized model. Spectroscopic analysis of the Au chains exhibit metallic character, irrespective of the twofold superstructure.
These findings are not compatible with a Peierls scenario, and instead provide evidence for a \textit{spin-ordered ground state} of the Si chains.
 
Experimentally, n-type Si(553) substrates (phosphorus doped) were cleaned by direct current heating (up to 1250~$^{\circ}$C). Au evaporation was performed on hot substrates ($\sim$650~$^{\circ}$C) followed by post annealing ($\sim$850~$^{\circ}$C) as in \cite{Crain2003}. This preparation method yields excellent long-range ordered nanowire arrays with Au and Si chains being separated by 1.48~nm each (see Fig. \ref{fig:Biasseries}), which is consistent previous experimental work \cite{Crain2003, Snijders2006}. All STM and STS measurements have been performed with an Omicron low-temperature STM. Spectroscopic data were obtained by current imaging tunneling spectroscopy (CITS) \cite{Hamers} which, simultaneously to a general constant current image, records a grid of I-V-curves on the same sample area. For the latter the feedback controller is turned off, resulting in a constant tip height while the tunneling voltage is ramped \cite{Hamers}.

Fig. \ref{fig:Biasseries} displays several constant current STM images as a function of tunneling bias (0.1~V to 1.4~V). Strikingly, this bias series of the unoccupied states exhibits \textit{both} reported periodicities of the Si chain ($\times$1 and $\times$3) at one fixed temperature (77~K). For low tunneling bias ($<$0.3~V) the Si chain features a $\times$1 periodicity, which for higher tunneling bias ($>$0.3~V) changes to a $\times$3 periodicity. This observation and especially the absence of a superstructure in the Si chain close to the Fermi level (low bias) is inconsistent with a simple CDW picture.

As an interesting detail, the transition from $\times$1 to $\times$3 periodicity does not occur at the same energy position for all chains. The two left Si chains display the transition to $\times$3 periodicity at lower bias (0.3-0.5~V) than the right chain does (0.4-0.6~V). A possible explanation for this behavior could be local defect-induced doping (see white circle in Fig. 1) as recently reported \cite{Barke2012} for the quasi-1D system Si(111)-(5$\times$2)-Au.

The second chain type, the Au chains, located in-between the Si chains, can easily be identified for low tunneling bias ($\leq$0.3~V) and show the reported twofold periodicity \cite{Ahn2005, Snijders2006}. For higher tunneling bias ($>$0.3~V) the Au contribution fades out and instead STM images are dominated by the threefold periodicity of the Si chains (see Fig. \ref{fig:Biasseries}). 

\begin{figure}[t!]
\includegraphics[width=\linewidth]{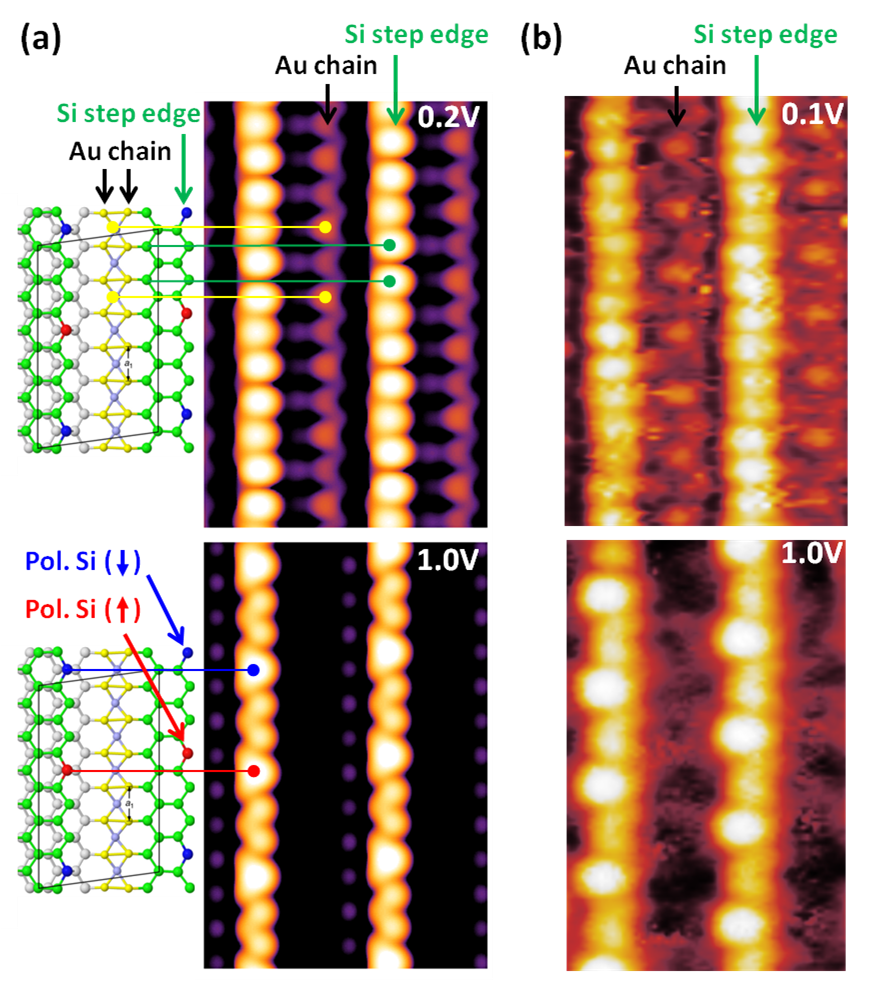}
 \caption{(a) Low and high bias simulated constant current STM image based on the spin-polarized DFT ground state.  A top view of the corresponding structural model is pictured on the left. The colored arrows allocate  different atom types (yellow: Au atoms; green: unpol. Si atoms; red and blue: spin-pol. Si atoms (spin-up and spin-down) to the simulated STM images. (b) Low bias (0.1~V) and high bias (1.0~V) experimental STM image.}
\label{fig:ExTheo}
\end{figure}

As an alternative to the Peierls scenario, the observed periodicities and their bias dependence can be explained by a model in which every third Si atom of the step edge is singled out by hosting a fully spin-polarized electron \cite{Erwin}. \color{black} Fig. \ref{fig:ExTheo}(a) displays two simulated STM images, one for low and one for high tunneling bias, based on the spin-polarized ground state from DFT \cite{Erwin}. The top view of the corresponding structural model is pictured to the left. The connecting lines allocate different atom types to their corresponding positions in the simulated STM images. 

The low bias image in Fig.~\ref{fig:ExTheo}(a) contains bright Si chain atoms showing a $\times$1 periodicity and clearly visible $\times$2 ordered Au contributions in-between. In contrast, the high bias picture (Fig. \ref{fig:ExTheo}(a) bottom) is dominated by the spin-polarized Si atoms resulting in threefold periodicity for the Si chain. The Au chains have almost lost their intensity. The $\times$2 periodicity of the Au chain visible at low bias originates from a small, alternating rotation of the Au dimers comprising the noble metal chain, which is a structural characteristic of the DFT ground state. Note that even though the Au chain consists of Au dimers, the simulated STM images reflect this only as a single Au row.

The experimental low and high bias STM images of Fig. \ref{fig:ExTheo}(b) are in excellent agreement with the corresponding simulated images:
the Si chains reflect the change in periodicity from $\times$1 to $\times$3 for increasing bias, while the Au chains disappear 
in both experiment and DFT simulation. Notably, even the shape of the experimentally observed Si atoms reveals a striking match with the theoretical image. 
In both corresponding high bias pictures, every third Si atom shows a bright mesa structure, while the less pronounced Si atoms between them exhibit a V-like shape. The in-line phase shift between the two Si chains visible in Fig. \ref{fig:ExTheo} (b) bottom is 3/2~a$_{0}$, where a$_{0}$ is the distance between two Si atoms along the step edge. This is consistent with the structural model (see Supplementary Figure S1 for a phase analysis).

In DFT the bias-dependent transition from $\times$1 to $\times$3 periodicity is caused by a prominent peak in the local DOS (LDOS) $\sim$0.5~eV above the Fermi level (E$_{\text{F}}$), which originates from the spin-polarized Si atoms (see Fig. 3a in \cite{Erwin}).
Indeed, if the spin polarization is artificially constrained to be zero in the calculation, then all Si step-edge atoms have identical unoccupied LDOS, which moreover reaches only $\sim$0.1~eV above the Fermi level \cite{Erwin}. Thus our observations strongly support the \textit{magnetic origin} of the LDOS at 0.5~eV. Thus, and due to the excellent match of our experimental data with DFT, one can conclude that a spin order scenario for the Si honeycomb chains is quite likely.

In addition, the calculated STM images also match the experimentally observed periodicity and bias dependence of the Au chain. At low bias, both images of Fig. \ref{fig:ExTheo} display a $\times$2 periodicity of the Au chain, while at high bias, the Au chain signal has nearly vanished.

\begin{figure}[b]
\includegraphics[width=\linewidth]{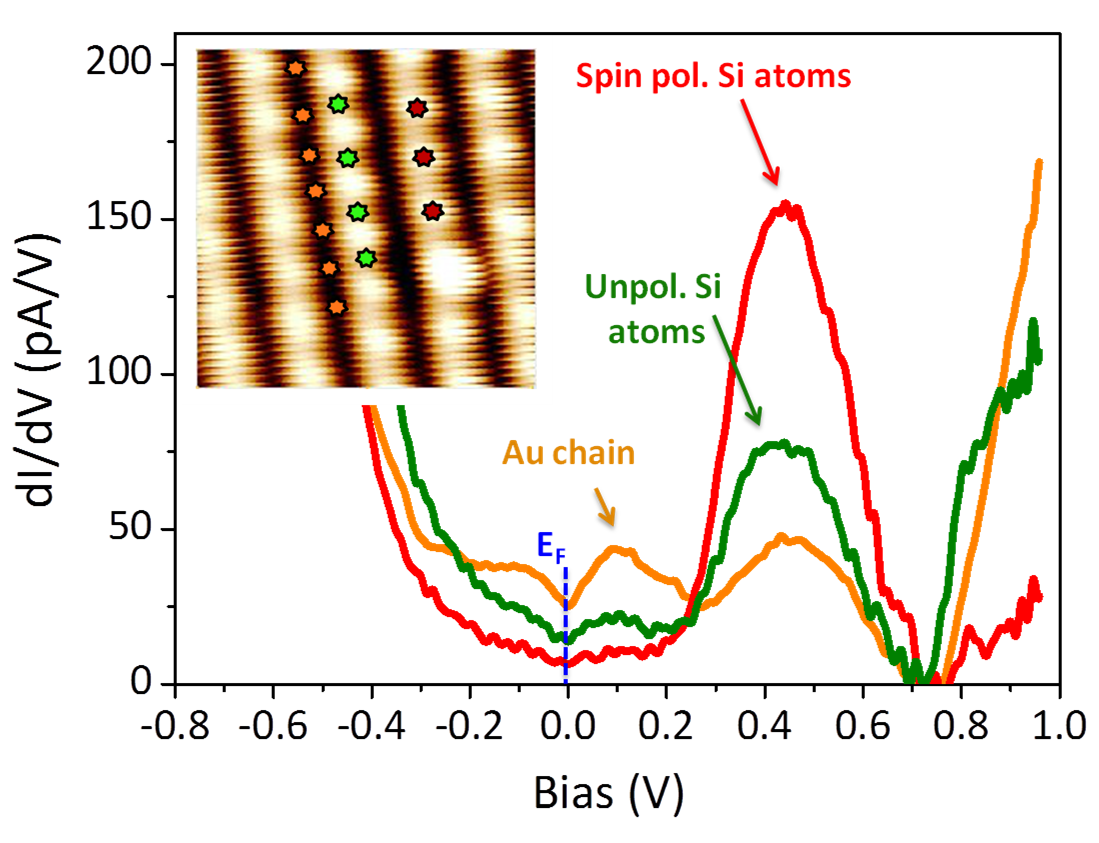}
 \caption{Local tunneling spectroscopy on individual atom positions. Inset: Constant current STM image (at the set point $U$~=~1.0~V, $I$~=~50~pA) recorded simultaneously with a grid of I-V-Spectra of the same sample area. The horizontal streaks are due to piezo hysteresis during spectrum acquisition. Main image shows three different dI/dV spectra. Orange curve: Spectra recorded on the Au chains e. g. as marked with the orange stars in the inset. Green and red curve represent spectra taken at unpolarized (green stars) and spin-polarized (red stars) Si atoms, respectively. Finite dI/dV signal of the Au chain at E$_{F}$ excludes a Peierls gap. The intense peak at $\sim$0.44~V can be assigned to the spin-polarized Si atoms.}
\label{fig:LocalSTS}
\end{figure}

\begin{figure}[b!]
\includegraphics[width=\linewidth]{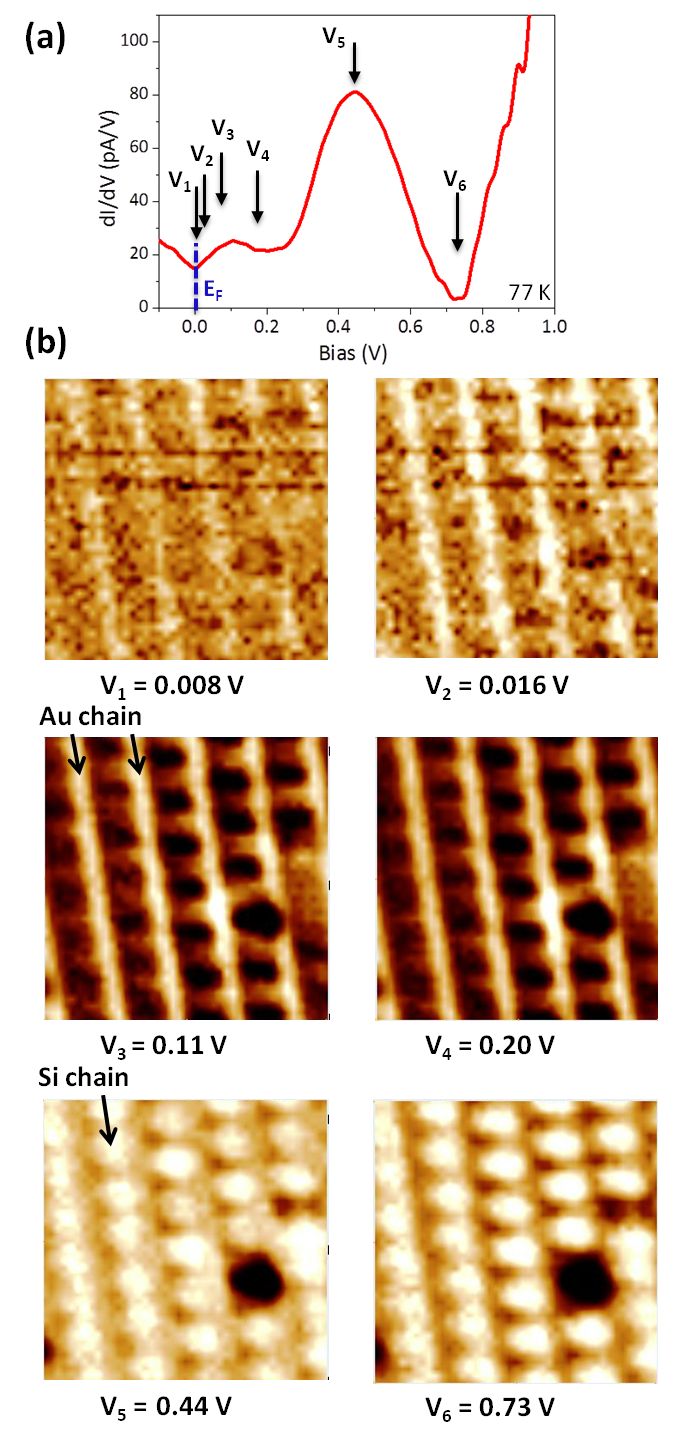}
 \caption{(a) dI/dV spectrum averaged over the area displayed by the simultaneously recorded constant current image shown in the inset of Fig. \ref{fig:LocalSTS} (set point $U$~=~1.0~V, $I$~=~50~pA). (b) Current maps at various tunneling bias for these tunneling conditions. Close to the Fermi level (bias V$_{1}$ and V$_{2}$) the current maps exhibit bright filaments, proving the metallic character of the Au chains. At bias V$_{5}$ and V$_{6}$ every third Si atom displays highly enhanced intensity, which can be directly related to the prominent peak in the LDOS at 0.44~V.}
\label{fig:CurrentMap}
\end{figure}
A more detailed analysis of the electronic properties of these different chain types can be achieved by local STS on the individual chains (see Fig. \ref{fig:LocalSTS}). For this purpose, we performed a grid of I-V-curves over a 7~nm $\times$ 7~nm sample area, while simultaneously recording a topography image of the same area, see inset of Fig. \ref{fig:LocalSTS}. 
The three dI/dV spectra displayed in Fig. \ref{fig:LocalSTS} represent the three different characteristic features of this system. The orange curve contains only spectra taken at the position of the Au chains exemplarily marked with the orange stars in the inset of Fig. \ref{fig:LocalSTS}. The green curve includes spectra recorded between the bright protrusions, i.e., at the position of the \textit{unpolarized} Si atoms, while the red spectrum represents the LDOS of the bright mesa atoms, i.e., the \textit{spin-polarized} Si atoms.

Since the Au chains show a finite LDOS at the Fermi level, one can safely exclude a Peierls gap. The spectra have been recorded at 77~K, i.e., far below the hypothetical Peierls transition temperature of the Au chains, which has earlier been estimated to be above room temperature (RT) \cite{Ahn2005}. 
Taken together with the close match with the DFT calculations one can dismiss a Peierls scenario for the Au chain, while the $\times$2 periodicity has to be explained by distorted Au dimers, which are an inherent structural feature.

Additionally, spectra taken at the mesa segments exhibit an intense peak at $\sim$0.44~V, again providing a direct match with the predicted spin order scenario. 

A corresponding feature in the unoccupied energy range is also visible in two-photon photoemission experiments \cite{Biedermann}, which, however, is representing an average over all atoms in both chains.
Subsequently, an increased DOS signal at every third Si atom has been reported by a STS study \cite{Snijders2012}. In contrast to \cite{Snijders2012}, our data do not exhibit a second strong peak closer to the Fermi level ($\sim$0.2~V) for the spin-polarized Si atoms, for which the spin model provides no match. Instead, we find that the dI/dV signal of the intense peak ($\sim$0.44~V) is more than an order of magnitude higher than in the low bias region ($\sim$0.2~V).

In order to map the LDOS of both chain types, we used CITS maps. Such "current maps" simply display the tunneling current at a designated tunneling voltage, i.e., the integrated LDOS. 
Fig. \ref{fig:CurrentMap}(a) shows a dI/dV spectrum, but in contrast to Fig. \ref{fig:LocalSTS} averaged over the scanned sample area. The black arrows mark the bias used in the current maps displayed in Fig. \ref{fig:CurrentMap}(b). The maps were not only taken at bias voltages corresponding to LDOS peak maxima, but also above the peak energies (i.e., at local LDOS minima), in order to obtain the full contrast of the integrated peak. Close to the Fermi level (bias 8 and 16~mV) the current maps exhibit bright lines. 
A comparison with the simultaneously recorded topographic image (Fig. \ref{fig:LocalSTS} inset) reveals that these lines originate from the Au chains.
This is indicative of \textit{metallic behavior}, and again rules out a Peierls-like band gap for the Au chains. For intermediate bias values (bias 0.11 and 0.22~V), the Au filaments get brighter, and also rungs between the filaments appear. 
At 0.44 and 0.73~V (i.e., at and beyond the LDOS peak) former dark areas (of bias 0.11 and 0.22~V) turn extremely bright, even brighter than the Au filaments. The distance between the bright protrusions corresponds to three times the Si edge atom distances $a_{0}$. Additionally, a comparison with the inset of Fig. \ref{fig:LocalSTS} reveals that the bright protrusions match the position of bright Si atoms. This confirms the result from local spectroscopy that the prominent peak in the LDOS relates to exactly every third Si step edge atom.

Based on these compelling observations and their comprehensive consistency with the DFT-based model of Ref. \cite{Erwin}, the physical mechanism underlying the higher order periodicity is reinterpreted as driven not by charge ordering, but by \textit{spin ordering}. As a consequence, the results imply that the system - contrary to previous perception - does not show a Peierls-type phase transition. Therefore, given the known absence of superstructures on the Si chains at RT \cite{Crain2003}, it will be interesting to explore the dynamics of the spin-polarized Si atoms as a function of temperature. This is inspired by recent molecular dynamics and kinetic Monte Carlo simulations suggesting a continuous crossover to a disordered situation at high temperatures \cite{Erwin2013}. Moreover, due to the subtle energetics of the spin alignment in the DFT model (supplement to \cite{Erwin}), it should be possible to intentionally modify the spin pattern by additional dopant atoms or magnetic impurities.

%
%
%


\subsection{}
\subsubsection{}

\begin{acknowledgments}
We are grateful to C. Blumenstein for technical advice, and acknowledge financial support by the Deutsche Forschungsgemeinschaft (through FOR1162 and FOR1700). Funding for this project was also provided by the Office of Naval Research through the Naval Research Laboratory's Basic Research Program. Computations were performed at the DoD Major Shared Resource Centers at AFRL and ERDC.
\end{acknowledgments}

%

\end{document}